\documentclass[aps,prl,twocolumn,superscriptaddress,showpacs,footnotebib]{revtex4-1}
\usepackage{amsmath}
\usepackage{color}
\usepackage{graphicx}
\usepackage{epstopdf}
\usepackage{epsfig}
\usepackage{amsfonts}
\usepackage[naturalnames]{hyperref}
\usepackage{verbatim}
\usepackage{tabularx}

\begin{document}
\graphicspath{{figure/}}

\title{Disorder-induced half-integer quantized conductance plateau in quantum anomalous Hall insulator-superconductor structures}
\author{Yingyi Huang}
\affiliation{Condensed Matter Theory Center and Joint Quantum
Institute, Department of Physics, University of Maryland, College
Park, Maryland 20742, USA}
\affiliation{State Key Laboratory of Optoelectronic Materials and Technologies, School of Physics, Sun Yat-sen University, Guangzhou 510275, China}
\author{F.~Setiawan}
\affiliation{Condensed Matter Theory Center and Joint Quantum
Institute, Department of Physics, University of Maryland, College
Park, Maryland 20742, USA}
\author{Jay D.~Sau}
\affiliation{Condensed Matter Theory Center and Joint Quantum
Institute, Department of Physics, University of Maryland, College
Park, Maryland 20742, USA}
\date{\today}
\begin{abstract}
A weak superconducting proximity effect in the vicinity of the topological transition of a quantum anomalous Hall system has been proposed as a venue to realize a topological superconductor (TSC) with chiral Majorana edge modes (CMEMs). A recent experiment [\href{http://science.sciencemag.org/content/357/6348/294} {Science \textbf{357},
294 (2017)}] claimed to have observed such CMEMs in the form of a half-integer quantized conductance plateau in the two-terminal transport measurement of a quantum anomalous Hall-superconductor junction. Although the presence of a superconducting proximity effect generically splits the quantum Hall transition into two phase transitions with a gapped TSC in between, in this Rapid Communication we propose that a
nearly flat conductance plateau, similar to that expected from CMEMs, can also arise from the percolation of quantum Hall edges well before the onset of the TSC or at temperatures much above the TSC gap.
 Our Rapid Communication, therefore, suggests that, in order to confirm the TSC, it is necessary to supplement the observation of the half-quantized conductance plateau with a hard superconducting gap (which is unlikely for a disordered system) from the conductance measurements or the heat transport measurement of the transport gap. Alternatively, the half-quantized thermal conductance would also serve as a smoking-gun signature of the TSC.
\end{abstract}

\maketitle

Recent years have seen a burgeoning interest in realizing topological superconductors (TSCs) which host zero-energy Majorana modes. These Majorana zero modes hold potential applications for a fault-tolerant topological quantum computation~\cite{RevModPhys.80.1083} owing to their non-Abelian braiding statistics~\cite{Read-2000-Paired,Ivanov-2001-Non}. They can be found in the vortex core of a two-dimensional (2D) chiral TSC with an odd integer Chern number. Recent theoretical studies~\cite{Qi-2010-Chiral,PhysRevB.83.100512,PhysRevB.92.064520} proposed to realize this chiral TSC using a quantum anomalous Hall insulator (QAHI) in proximity to an $s$-wave superconductor (SC).

\begin{figure}[t!]
\centering
\includegraphics[width=\linewidth]{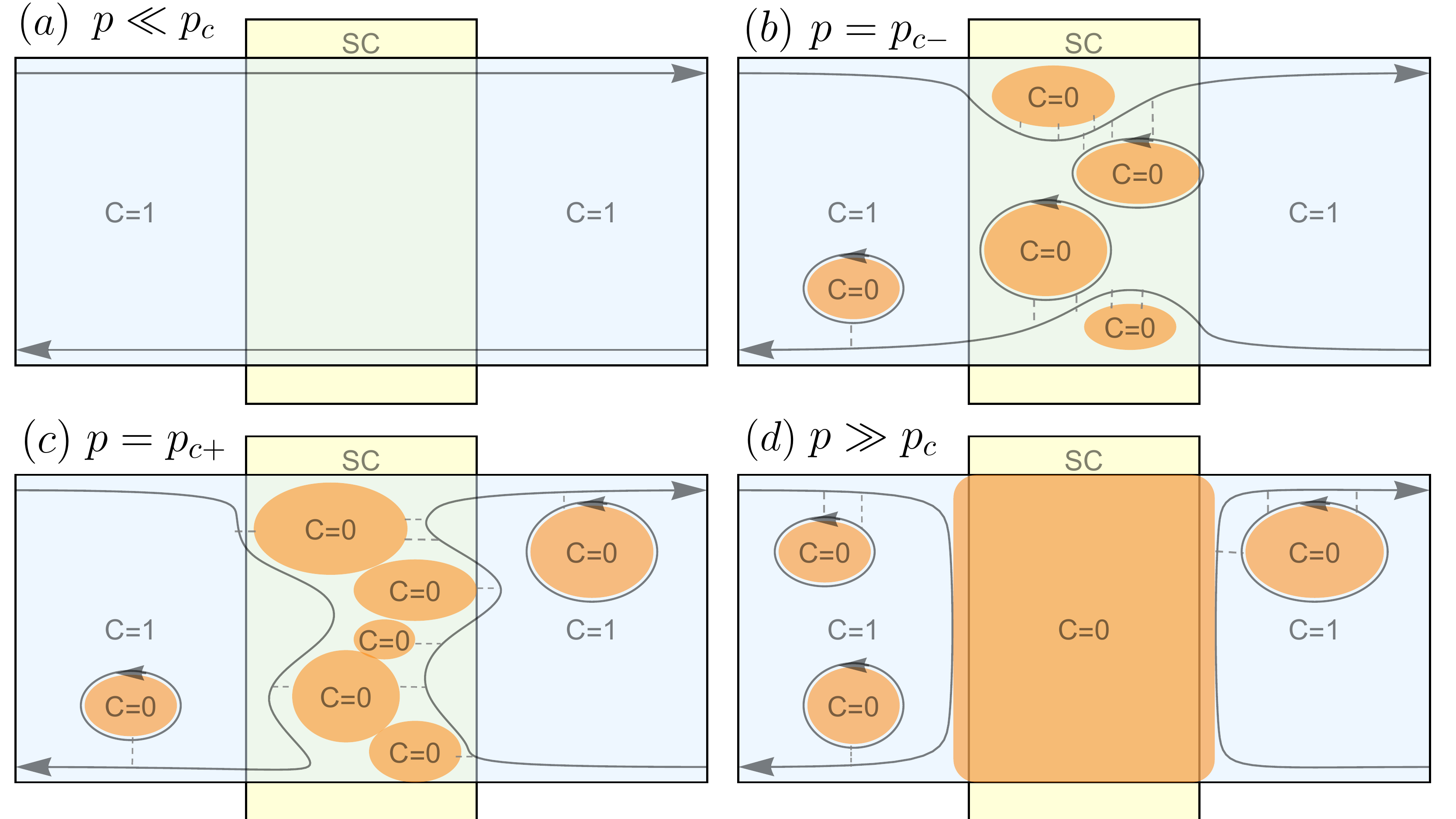}
\caption{\label{Fig1} Schematics of the magnetic-field-induced percolation in a disordered QAHI-SC-QAHI junction. The middle QAHI region is proximitized by an $s$-wave SC (the yellow rectangle). Four different percolation stages of trivial insulator phases (the orange region) with $\mathcal{C}=0$
 (corresponding to $\mathcal{N}=0$) and QAHI phases (the light-blue region) with $\mathcal{C}=1$ (corresponding to $\mathcal{N} = 2)$. We consider a strongly disordered system where $\mathcal{N} =1$ domains do not form. The four stages are characterized by $p$, the proportion of the $\mathcal{C}=0$ phase, which changes with the magnetic field. (a) In the strong magnetic field regime where $p$ is far below the percolation threshold ($p\ll p_c$), the system is in the $\mathcal{C}=1$ phase. The edge states (the arrowed lines) are perfectly transmitted across the junction. (b) During the magnetization reversal, the $\mathcal{C} =0$ phase domains grow. The edge states wind around the domains in the SC region and leak into adjacent chiral loops (the dashed lines). (c) When $p$ is slightly above the percolation threshold ($p=p_{c+}$), the domains are connected across the junction width and the edge states can no longer be transmitted across the junction. (d) When $p\gg p_c$, the edge states are normally reflected by the $\mathcal{C}=0$ phase outside the SC region.}
\end{figure}

The quantum anomalous Hall (QAH) state is a quantum Hall (QH) state without an external magnetic field which can be realized in a 2D thin film of a magnetic topological insulator (TI) with ferromagnetic ordering~\cite{2006-Qi-Topological,2008-Liu-Quantum,Liu-2016-The,yu2010quantized,chang2013experimental}. For the regime where the ferromagnetic-induced exchange field strength $|\lambda|$ is greater than the hybridization gap  $|m_0|$ induced by the coupling between the top and bottom surfaces, the system has a Chern number of $\mathcal{C} = \lambda /|\lambda|$ and in the opposite limit where $|\lambda| < |m_0|$, $\mathcal{C} = 0$~\cite{PhysRevB.92.064520,2014-Wang-Universal}. By changing the applied magnetic field over a relatively small range, a topological phase transition can be induced between the QAHI with $\mathcal{C}=1$ and the trivial insulator state with $\mathcal{C}=0$~\cite{Chang-2016-Observation}. When the QAH is proximitized by an $s$-wave SC, the $\mathcal{C} = 1$ and $\mathcal{C} = 0$ phases are driven into $\mathcal{N} = 2$ and $\mathcal{N} = 0$ phases~\cite{Qi-2010-Chiral}, respectively, where $\mathcal{N}$ denotes the number of  chiral Majorana edge mode (CMEM). At the transition between these two phases, there exists an $\mathcal{N} = 1$ gapped TSC~\cite{Qi-2010-Chiral,PhysRevB.83.100512}. Since a single CMEM carries one-half of the incoming charges, it manifests as a half-integer quantized $e^2/2h$ plateau in the conductance between two normal leads and an integer quantized $e^2/h$ peak in the conductance between a normal lead and the SC measured at the coercive field~\cite{PhysRevB.83.100512,PhysRevB.92.064520}. A recent experiment~\cite{experiment} observed these two transport signatures in a doped magnetic QAHI thin film proximitized by an $s$-wave SC. Although these transport signatures are consistent with the existence of an $\mathcal{N} = 1$ TSC with a single CMEM in a clean system, the disorder in the experimental system might significantly reduce the topological gap and phase space of the $\mathcal{N}=1$ TSC.

In this Rapid Communication, we show that the two proposed transport signatures for the $\mathcal{N} = 1$ phase can generically occur in a disordered QAHI-SC-QAHI junction even in phases where the CMEM is absent, such as in the $\mathcal{C}=1$ ($\mathcal{N}=2$) phase or in the $\mathcal{N} = 1$ TSC but at temperatures above the gap. We consider the disordered QAH system to be inhomogeneous with smoothly varying magnetization~\cite{chalker1988percolation} which leads to a network of domain walls between phases with different Chern numbers. Such domain walls have been invoked in Ref.~\cite{chen2016effects} to understand the Hall conductance in this system. Here, we consider the disorder strength to be stronger than the superconducting pairing potential such that there are no $\mathcal{N}=1$ domains in the system.

Figure~\ref{Fig1} shows the evolution of the domain-wall structure of the phases in the QAH system as the magnetic field is varied. In the limit of strong magnetic field, the system is in a single-domain $\mathcal{C} = 1$ phase [as shown in Fig.~\ref{Fig1}(a)] with a large average magnetization. In this regime, the edge states are perfectly transmitted across the junction. During the magnetization reversal, the proportion $p$ of the $\mathcal{C} = 0$ domain (domain with small average magnetization) increases [Fig.~\ref{Fig1}(b)]. Since the chiral edge states live at the boundary between the $\mathcal{C} = 0$ and the $\mathcal{C} = 1$ domains, the edge state has to wind around the $\mathcal{C} = 0$ domains which increases the electron trajectory length $L$ and hence the number of Andreev scatterings in the SC region. As $p$ approaches the percolation threshold $p_c$ (where the $\mathcal{C} = 0$ domains become connected into a cluster spanning across the junction width), $L \rightarrow \infty$. In addition, quasiparticles on the chiral edge can leak by quantum tunneling into adjacent chiral loops associated with the domains as shown in Fig.~\ref{Fig1}. These chiral loops can be assumed to be in equilibrium. As a result, at $p\approx p_c$, as we will show, the leakage of quasiparticles leads to eventual absorption of the initial quasiparticle for large lengths $L$, giving rise to a nearly flat $e^2/2h$ two-terminal conductance plateau.  As $p$ increases above $p_c$,  the edge states can no longer be transmitted across the junction. For $p \gg p_c$, the electrons undergo perfect normal reflections outside the SC region by the $\mathcal{C} = 0$ domain as shown in Fig.~\ref{Fig1}(d).

We describe the low-energy edge modes of the QAHI-SC structure by a one-dimensional Hamiltonian,
\begin{equation}
H=\frac{1}{2}\int dx\mathcal{C}^\dag(x) \mathcal{H}_{\mathrm{BdG}}(x)\mathcal{C}(x),
\end{equation}
where
\begin{equation}
\mathcal{H}_{\mathrm{BdG}}(x) = -iv\tau_0\partial_x-\mu(x)\tau_z+\frac{1}{2}\left\{-i\partial_x,\Delta(x)\tau_x\right\}
\label{model}
\end{equation}
is the Bogoliubov-de Gennes (BdG) Hamiltonian
and $\mathcal{C}(x)=[c(x),c^\dag(x)]^T$ is the Nambu spinor with $c(x)$ and $c^\dagger(x)$ being the electron annihilation and creation operators, respectively. Here, $v$ is the edge mode velocity, $\mu$ is the chemical potential, $\Delta$ is the effective $p$-wave pairing potential of the proximity-induced superconductivity, and $\tau_{x,y,z}$  are the Pauli matrices in the particle-hole space. For the QAHI region, we set $ \Delta = 0$, whereas for the SC region, we set $\mu(x)$ and $\Delta(x)$ to be spatially varying along the electron trajectory length $L$. For simplicity, we work in the units where the Planck constant $\hbar$, the Boltzmann constant $k_B$, and edge velocity $v$ are all set to $1$. We note that the term $\partial_x$ in the Hamiltonian comes with the anticommutation relation $\{,\}$ to ensure the Hermiticity of the Hamiltonian. The $p$-wave pairing amplitude $\Delta(x)$ is induced from the proximity effect of an $s$-wave SC with a pairing potential $\Delta_s(x)$. This cannot occur in a strictly spin polarized edge state. However, since the QAH system arises from a TI, which is a strongly spin-orbit-coupled system, we expect the spin-polarization of the chiral edge state to vary with momentum (similar to the spin texture in a TI~\cite{Hasan-2010-Colloquium} on a scale of the spin-orbit length $k_{\mathrm{so}}^{-1}$ where $k_{\mathrm{so}}$ is related to the exchange field $\lambda$ by $k_{\mathrm{so}}\sim \lambda/v$). Within this model, $\Delta(x)\sim v\Delta_s(x)/\lambda$ (see the Supplemental Material~\cite{suppl1} for the derivation).

\begin{figure}
\centering
\includegraphics[width=\linewidth]{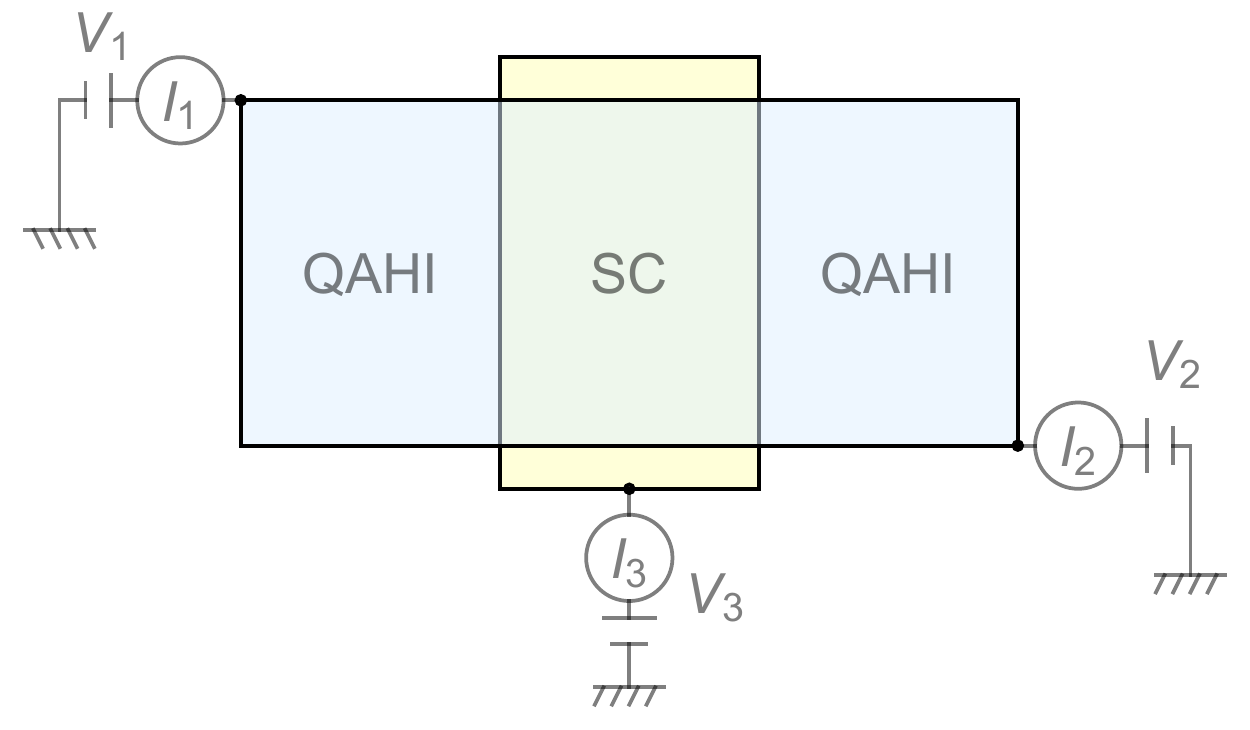}
\caption{\label{Fig2} Schematic of the setup used to measure the conductance in a QAHI-SC-QAHI junction. To measure $G_{12}$, we consider the SC to be floating (i.e., $I_3 = 0$) and the voltages $V_1$ and $V_2$ to be applied to leads 1 and 2, respectively. For the case where $G_{13}$ is measured, the SC is grounded (i.e., $I_3 \neq 0$), lead 2 is removed, and the voltages $V_{1}$ and $V_{3}$ are applied to lead 1 and the SC, respectively.}
\end{figure}

The conductance of the three-terminal junction shown in Fig.~\ref{Fig2} where terminal 3 is connected to the SC can be computed from Bogoliubov quasiparticle transmission and reflection probabilities using a generalized Landauer-B\"{u}ttiker formalism~\cite{Datta,2008-Entin-Conductance,takane1992conductance}. Using this formalism the currents $I_{1,2}$ shown in Fig.~\ref{Fig2} are found to be
\begin{subequations}\label{Landauer}
\begin{align}
I_1&=\frac{e^2}{h}[(1-g_{11})(V_1-V_3)-g_{12}(V_2-V_3)], \label{Landauera}\\
I_2&=\frac{e^2}{h}[-g_{21}(V_1-V_3)+(1-g_{22})(V_2-V_3)],\label{Landauerb}
\end{align}
\end{subequations}
where $V_1$ and $V_2$ are the voltages of leads $1$ and $2$, respectively, $V_3$ is the voltage of the SC
and $g_{ij}$ are effective dimensionless conductances from lead $i$ to lead $j$ due to the chiral edges.
Experimentally, the conductance is measured using a two-terminal setup, i.e., the setup in Fig.~\ref{Fig2} with either the current $I_2 = 0$ (grounding) or $I_3 = 0$ (floating SC case) depending on the measured transport properties.
For the case of floating SC, we obtain the conductance between leads 1 and 2 from Eq.~\eqref{Landauer} and the current conservation equation ($I_1+I_2=0$) as
\begin{equation}
G_{12}\equiv\frac{I_1}{V_1-V_2}=\frac{e^2}{h}\left[\frac{g_{21}g_{12}-(1-g_{11})(1-g_{22})}{g_{12}+g_{21}+g_{11}+g_{22}-2}\right].
\label{G12}
\end{equation}
For the case where the SC is grounded, lead 2 is removed ($I_2=0$), and the conductance between lead $1$ and the SC can be obtained from Eq.~\eqref{Landauer} to be
\begin{equation}
G_{13}\equiv\frac{I_1}{V_1-V_{3}}=\frac{e^2}{h}\left[\frac{(1-g_{11})(1-g_{22})-g_{12}g_{21}}{1-g_{22}}\right].
\label{G13}
\end{equation}

To compute the parameters $g_{ij}$ that determine the measured conductances [Eqs.~\eqref{G12} and~\eqref{G13}], we need to
consider a microscopic model of the chiral edges in the vicinity of the SC.
For  $p < p_c$,  we assume that $g_{11}=g_{22} = 0$  as the chiral edge state emanating from $I_1$ can only be transmitted to $I_2$, whereas for $p > p_c$, $g_{12}=g_{21}=0$ as the edge states can only undergo reflection. The above condition holds in the typical case where the width of the system is larger than the correlation length at some finite distance away from the critical point such that the edge states (as shown in Fig.~\ref{Fig1}) do not couple to each other. For computational simplicity, we assume that the conductances are the same for the left- and right-incoming modes, i.e., $g_{12} = g_{21}$ and $g_{11} = g_{22}$ which is true for a symmetric junction. Our results, however, hold in general and do not qualitatively depend on this assumption.

The microscopic values of the parameter $g_{11}$ or $g_{12}$ (whichever is nonvanishing) are determined by a combination of superconductivity and dephasing. Without superconductivity, $g_{11}=1$, which results in a QH transition seen between the two
quantized values of $G_{12}=1$ to $G_{12}=0$ with no intervening plateau. The introduction of superconductivity on a
disordered chiral edge allows for Andreev scattering which gives rise to an intervening plateau. However, to obtain an intervening plateau that is stable at
low temperatures one must account for dephasing through tunneling from the chiral edge into the disjointed chiral loops $L_n$ (seen in Fig.~\ref{Fig1}).
The nonvanishing conductance $g_{11}$ or $g_{12}$ (depending on whether $p>p_c$ or not) is determined by the transconductance $g_{\mathrm{trans}}$
across the incoherent chiral edge (coupled to an SC) that results from the tunneling into the loops $L_n$. To determine $g_{\mathrm{trans}}$ sufficiently close to
the percolation point, where the loops $L_n$ are expected to be larger than the finite-temperature and interaction-induced dephasing length $v\tau_\varphi$ (where $v$ is the chiral edge velocity and $\tau_\varphi$ is the dephasing time),
we  assume the loop $L_n$ to be a reservoir in equilibrium at voltage $v_n$ (relative to the SC). Furthermore, we assume that the coupling between
the loop $L_n$ and the SC is weak enough to allow incoherent transfer of Cooper pairs through a resistance $R_n \sim v\{L \tau_{\varphi}[\Delta_{s}(x_n)]^2\}^{-1}$ between
them. To understand the origin of the resistance $R_n$, we consider tunneling between the chiral loop and the SC which
leads to a conductance $G_S\sim n_{\mathrm{ch}}G_N^2$~\cite{Beenakker1992Quantum} where $n_{\mathrm{ch}}$ is the number of low-energy states (energy range of $\sim\tau_\varphi^{-1}$) in the chiral loop. The proximity gap $\Delta_s\propto G_N$ is proportional to the normal-state conductance $G_N$
per channel.
Given the voltages $v_n$ and the voltage difference $V_{\mathrm{in}}-V_3$ between the incoming edge and the SC, the transconductance is
given by~\cite{suppl2}
\begin{align}
&g_{\mathrm{trans}}=\Lambda_{\mathrm{in},\mathrm{out}}+\sum_n \Lambda_{n,\mathrm{out}} \frac{v_n}{V_{\mathrm{in}}-V_3},\label{eqgch}
\end{align}
where $\Lambda_{\mathrm{in},\mathrm{out}}$ and $\Lambda_{n,\mathrm{out}}$ are conductances obtained from the multiterminal Landauer-B\"{u}ttiker formalism~\cite{datta1997electronic}.
Specifically, the incoherent chiral edge may be thought of as a multiterminal system with leads at the in and out ends as well as each
of the loops $L_n$. We can then define the response of the current in lead $n$ to the
voltage in lead $m$ by
\begin{equation}\label{eq:deltaS}
 \Lambda_{mn}=\int_{-\infty}^{\infty} dE \left(-\frac{\partial f_T(E)}{\partial E}\right )(|t^N_{mn}(E)|^2-|t^A_{mn}(E)|^2),
\end{equation}
where $f_T(E) = 1/(e^{E/T}+1)$ is the Fermi distribution, $t^N_{mn}(E)$ and $t^A_{mn}(E)$ are the normal and Andreev scattering amplitudes at energy $E$, respectively, from the lead $m$ into the lead $n$. Given $\Lambda_{mn}$,
the voltages $v_n$, that appear in Eq.~(\ref{eqgch}), can be determined
recursively as one follows the loops down the chiral edge which are given by
\begin{align}
&\frac{v_n}{V_{\mathrm{in}}-V_3}=\frac{\Lambda_{\mathrm{in},n}+\sum_{m<n}\Lambda_{mn}\frac{v_m}{V_{\mathrm{in}}-V_3}}{R_n^{-1}+\Lambda_{n,\mathrm{out}}+\sum_{m>n}\Lambda_{nm}}.
\end{align}
These relations as well as Eq.~(\ref{eqgch}) can be derived from the current conservation equation at each loop as detailed in the Supplemental Material~\cite{suppl2}.

The scattering amplitudes $t^{N,A}_{mn}$ are the components of $2 \times 2$ transmission (along the chiral edge) matrices acting in the particle-hole basis which is given by
\begin{equation}
\mathcal{T}_{mn}(E)=\begin{pmatrix} t^N_{mn}(E) & t^{A*}_{mn}(-E) \\ t^A_{mn}(E) & t^{N*}_{mn}(-E) \end{pmatrix}.
\end{equation}
The inhomogeneity of the chemical potential and pairing potential along the loop is accounted by matching the incoming and outgoing edge modes in the SC region with spatially varying $\mu(x)$ and $\Delta(x)$, where (see the Supplemental Material~\cite{suppl3} for the derivation)
\begin{equation}\label{eq:scatmat}
\mathcal{T}_{mn}(E) =\zeta_{mn} \prod_{m<j<n} e^{i\tilde{v}_j^{-1/2}(\mu_j\tau_z+E\tau_0)\tilde{v}_j^{-1/2}\ell},
\end{equation}
with $\tilde{v}_j=v\tau_0+\Delta_j\tau_x$ being the effective edge mode velocity at lattice site $j$ and $\ell$ being the lattice constant.
Here $\zeta_{mn}=\Omega_m\Omega_n\prod_{m<j<n}(1-\Omega_j^2)^{1/2}$ is a numerical factor that is related to the
couplings $|\Omega_j|<1$ of the chiral edge to the lead $j$ ($\Omega_{\mathrm{in}}=\Omega_{\mathrm{out}}\equiv 1$).

\begin{figure}[h!]
\centering
\includegraphics[width=0.8\linewidth]{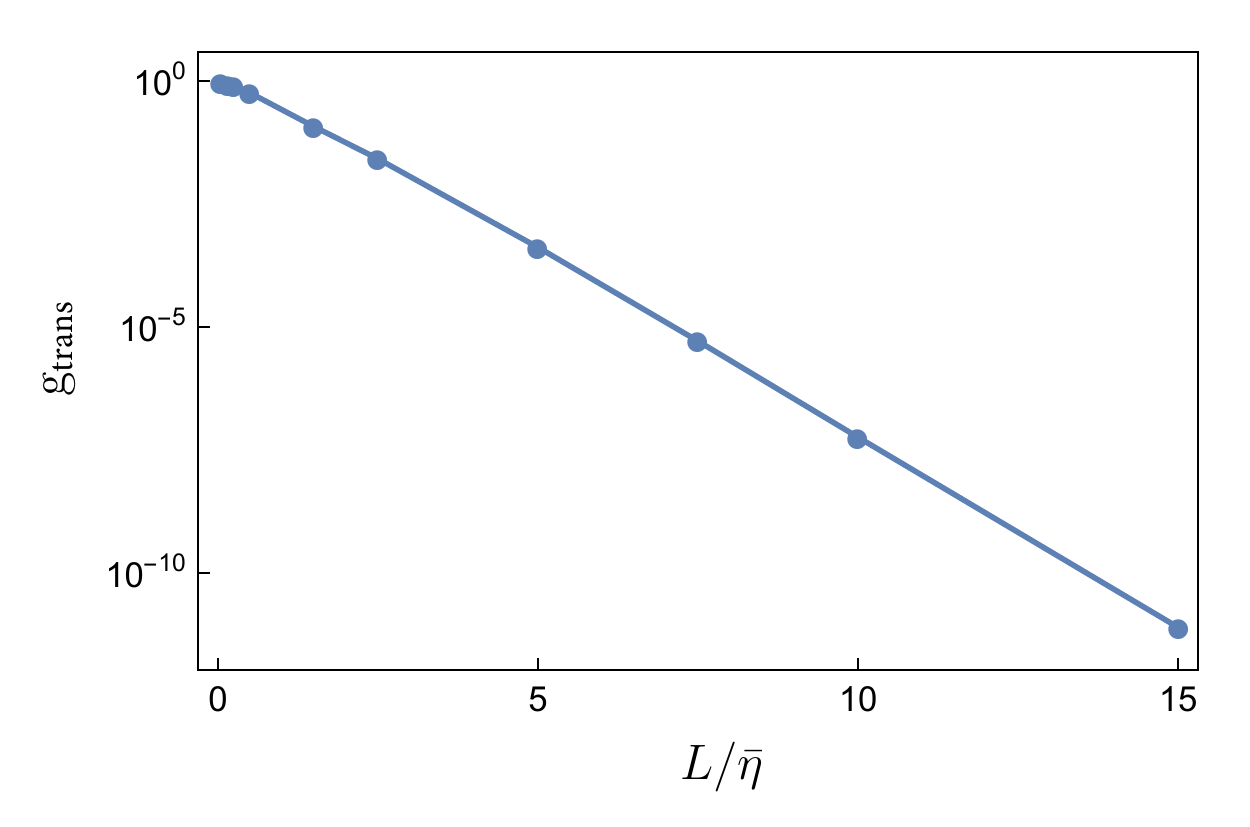}
\caption{\label{Fig3} Semilogarithmic plot of zero-energy effective transconductance $g_{\mathrm{trans}}$ vs electron trajectory length $L/\overline{\eta}$ where $\overline{\eta}= v/\overline{\Delta}$ is the average dimensionless $p$-wave superconducting coherence length. We consider spatially varying $\Delta(x)$ and $\mu(x)$ where the values of $\Delta(x)\in [0,0.1]$ and $\mu(x)\in [-0.01,0.01]$ are drawn from uniform distributions.
Note that $g_{\mathrm{trans}}$ exponentially decays with $L$. For $p < p_c$, $g_{\mathrm{trans}} =  g_{12} = g_{21}$ and $g_{11} = g_{22} = 0$, whereas for $p > p_c$, $g_{\mathrm{trans}} = g_{11} = g_{22}$ and $g_{12} = g_{21} = 0$.  Parameters used are edge mode velocity $v=1$, temperature $T=0.01$, resistance $R(x)=0.1/\{L[\Delta (x)]^2\}$, and coupling between the edge state and  loop $\Omega(x)=0.3$ for all $x$'s.}
\end{figure}

 From Eq.~\eqref{eq:scatmat}, we calculate the zero-bias net scattering probability $\Lambda_{mn}$ [Eq.~\eqref{eq:deltaS}] which is then used to compute the transconductance $g_{\mathrm{trans}}$ of a chiral edge [Eq.~\eqref{eqgch}], which is ultimately used to compute the two-terminal conductance [Eqs.~\eqref{G12} and~\eqref{G13}]. Figure~\ref{Fig3} shows the calculated $g_{\mathrm{trans}}$ as a function of the electron trajectory length $L/\overline{\eta}$ where $\overline{\eta} = v k_{\mathrm{so}}/\overline{\Delta}_s$ is the mean value of the dimensionless $p$-wave superconducting coherence length with $k_{\mathrm{so}}\sim$ (50 nm)$^{-1}$. From Fig.~\ref{Fig3}, we can see that $g_{\mathrm{trans}}$ decays exponentially with $L$. The electron trajectory length $L$ increases as the proportion $p \rightarrow p_c$ where the percolation threshold $p_c$ corresponds to the magnetic field near the coercive field. Near $p_c$, $L$ obeys the scaling relation~\cite{RevModPhys.64.961},

\begin{equation}
L=L_0|p-p_c|^{-\nu d_h}=L_0|p-p_c|^{-(1+\nu)},
\label{scaling}
\end{equation}
where for the 2D case considered here, the correlation length exponent $\nu$ is $\frac{4}{3}$~\cite{nu} and the fractal dimension of the hull $d_h$ is $(1+\nu)/\nu$~\cite{dh}. Using Eqs.~\eqref{eqgch}-\eqref{scaling}, we have $g_{\mathrm{trans}} \rightarrow 0$ as $p \rightarrow p_c$.

\begin{figure}
\centering
\includegraphics[width=0.8\linewidth]{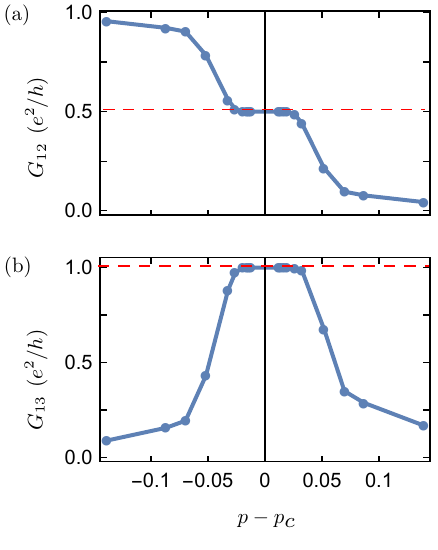}
\caption{Conductances (a) $G_{12}$ and (b) $G_{13}$ as a function of $p$ near the percolation threshold $p_c$. $G_{12}$ exhibits a half-integer quantized plateau at $p = p_c$, whereas $G_{13}$ shows an integer quantized peak at $p = p_c$.  The red dashed lines denote (a) $G = e^2/2h$ and (b) $G = e^2/h$. We set $L_0$ in Eq.~\eqref{scaling} to be $\overline{\eta}/2000$ so that the conductance plateaus have a short width near $p_c$. The parameters used here are the same as those used in Fig.~\ref{Fig3}. The plateau width is stable at relatively low temperatures where the plateau width does not change in the low-temperature regime.}\label{Fig4}
\end{figure}

Next, we computed $G_{12}$ and $G_{13}$ using Eqs.~\eqref{G12} and ~\eqref{G13}, respectively, for a specific disorder realization. Figure~\ref{Fig4} shows the numerically calculated $G_{12}$ and $G_{13}$ as functions of $p-p_c$ near the percolation threshold $p_c$. As seen from the plot, the conductance $G_{12} \simeq e^2/h$ for $p < p_c$ and $G_{12} \simeq 0$ for $p > p_c$ with an exponentially flat $e^2/2h$  plateau at $p_c$  whereas the conductance $G_{13} \simeq 0$ for $p < p_c$ and $p > p_c$ with an $e^2/h$ peak at $p_c$.
Close to $p_c$, we can write $G_{12}$ and $G_{13}$ by using Eqs.~\eqref{G12}-\eqref{scaling} as
\begin{align}
G_{12}\approx
\begin{cases}
\frac{e^2}{2h}(1+e^{-2\alpha|p-p_c|^{-(1+\nu)}}), & \text{for\hspace{0.3cm}} p = p_{c-}, \\
\frac{e^2}{2h}(1-e^{-2\alpha|p-p_c|^{-(1+\nu)}}), & \text{for\hspace{0.3cm}} p = p_{c+},
\end{cases}
\end{align}
and
\begin{align}
G_{13}\approx
\begin{cases}
\frac{e^2}{h}(1-e^{-4\alpha|p-p_c|^{-(1+\nu)}}), & \text{for\hspace{0.3cm}} p = p_{c-}, \\
\frac{e^2}{h}(1-e^{-2\alpha|p-p_c|^{-(1+\nu)}}), & \text{for\hspace{0.3cm}} p = p_{c+},
\end{cases}
\end{align}
where $\alpha$ is the inverse length scale for the exponential decay of $g_{\mathrm{trans}}$. At $p = p_c$, $G_{12}$ and $G_{13}$ are perfectly quantized at $e^2/2h$ and $e^2/h$, respectively, with exponentially flat plateaus. These plateaus, which originate from the disorder effect, resemble the experimental data~\cite{experiment} claimed to be the signatures of CMEMs.
The width of the disorder-induced plateau decreases with decreasing pairing amplitude $\Delta$ as discussed in the Supplemental Material~\cite{suppl4}.

Our results, based on a classical percolation model for the QH transition, are valid at high temperatures where the chiral edge becomes effectively
long enough to produce the plateaus in Fig.~\ref{Fig4}. This classical percolation picture is a reasonable description
of the QH transition away from the critical point, at a relatively high temperature~\cite{chalker1988percolation},
 or in the presence of dephasing arising from the interplay of interaction, disorder, and temperature~\cite{Kapitulnik-2001-Effects}. Such dephasing requires the equilibration rate of quasiparticles in the loop being fast compared to tunneling as in our simple model of dephasing.  The equilibration rate goes to zero as $T \rightarrow 0$. However, for appropriate interaction strengths and pairing potentials, the condition of strong dephasing can be satisfied to arbitrarily low temperatures leading to a weakly temperature-dependent plateau at low temperatures~\cite{suppl5}. On the other hand, the plateau that arises from thermal fluctuations (without quasiparticle leakage between the chiral edge and the adjacent chiral loops) is strongly temperature dependent~\cite{suppl5}.

The $e^2/2h$ plateau shown in Fig.~\ref{Fig4} would describe results not only in phases other than the $\mathcal{N} = 1$ TSC, but also in the $\mathcal{N}=1$ phase for temperatures above the topological gap. At such high temperatures, $g_{\mathrm{trans}}$ would vanish because the edge quasiparticles could escape into the bulk by thermal excitations which makes it difficult to ascribe the conductance plateau to the topological properties of the TSC. Additionally, it has been proposed that, in the limit of strong disorder, the gapped $\mathcal{N}=1$ TSC may be replaced by a gapless Majorana metal phase even at zero temperature~\cite{Senthil-2000-Quasiparticle}, which may also produce an $e^2/2h$ plateau.

Although our results do not contradict the theoretical existence of the $\mathcal{N} = 1$ TSC phase  (which is likely although not inevitable) in the vicinity of the QAH transition,
the nearly quantized $e^2/2h$ conductance plateau observed in the recent experiment~\cite{experiment} cannot serve as an
experimental evidence for the $\mathcal{N} = 1$ TSC as it is likely to arise outside the TSC phase as well. In principle, observing the stabilization of the plateau to a more
perfectly quantized plateau as the temperature is lowered together with either a hard superconducting gap from the electrical conductance measurement (which is unlikely for a disordered system) or a thermal transport gap would be the signatures of an $\mathcal{N}=1$ TSC. Another smoking-gun signature is the half-quantized thermal conductivity $K_H = (\pi k_B)^2T/6h$~\cite{Kane-1997-Quantized} which would rule out the classical percolation-based model and the Majorana metal phase as they would have a large nonuniversal longitudinal thermal conductance.

\textit{Note added}. Upon completion of our Rapid Communication, we became aware of a recent paper~\cite{Wen} related to our work.

We thank K. T. Law for valuable discussions that drew our attention to Ref.~\cite{experiment}, which motivated this Rapid Communication. J.D.S. acknowledges stimulating discussions with S.-C. Zhang that motivated the discussion of the low-temperature limit. We also thank Y. Alavirad, C.-K. Chiu, and Q.-L. He for helpful discussions. This Rapid Communication was supported by  JQI-NSF-PFC, Sloan Research Fellowship and NSF-DMR-1555135 (CAREER). Y.H. is grateful to the China Scholarship Council for financial support. We acknowledge the University of Maryland supercomputing resources~\cite{hpcc} made available in conducting the research reported in this Rapid Communication.

\bibliography{QAHSC}

\onecolumngrid
\vspace{1cm}
\begin{center}
{\bf\large Supplemental Material}
\end{center}
\vspace{0.5cm}

\setcounter{secnumdepth}{3}
\setcounter{equation}{0}
\setcounter{figure}{0}
\renewcommand{\theequation}{S-\arabic{equation}}
\renewcommand{\thefigure}{S\arabic{figure}}
\renewcommand\figurename{Supplementary Figure}
\renewcommand\tablename{Supplementary Table}
\newcommand\Scite[1]{[S\citealp{#1}]}
\makeatletter \renewcommand\@biblabel[1]{[S#1]} \makeatother
\section{Derivation of effective \texorpdfstring{$p$}{p}-wave pairing potential from the minimal QAH two-band model}
In this appendix, we derive the $p$-wave superconducting pairing potential $\Delta$ from the minimal two-band model of the QAH where we consider the superconducting term as a perturbation in the Hamiltonian. The two band-model for the QAH~\cite{PhysRevB.92.064520} is given by $H=\sum_{\textbf{p}}\mathcal{C}_{\textbf{p}}^\dag\mathcal{H}_{\textbf{p}}\mathcal{C}_{\textbf{p}}$, with $\mathcal{C}_{\textbf{p}}=(c^t_{\textbf{p}\uparrow},c^t_{\textbf{p}\downarrow},c^b_{\textbf{p}\uparrow},c^b_{\textbf{p}\downarrow})^T$ and
\begin{equation}
\mathcal{H}({\textbf{p}})=-vp_x\sigma_y\zeta_z+vp_y\sigma_x\zeta_z+m_0\zeta_x+\lambda\sigma_z,
\label{microH}
\end{equation}
where $c^l_{\textbf{p}s}$ annihilates an electron with momentum $\textbf{p}$ and spin $s=\uparrow,\downarrow$ in the top or bottom surfaces $l=t,b$. Here, $p_{x,y}$ are the momentum operators, $\zeta_{x,y,z}$ and $\sigma_{x,y,z}$ are the Pauli matrices for the orbital and spin degree of freedom, respectively. In the above Hamiltonian, the first two terms are the spin-orbit coupling term in the TI, $\lambda$ is the ferromagnetic-induced exchange field strength and $m_0$ is the hybridization gap. For $|\lambda|>|m_0|$, the system has a Chern number $\mathcal{C} =\lambda/|\lambda|$ and for $|\lambda| <|m_0|$, $\mathcal{C} = 0$.

Since the gap between the upper bands ($|m_0| + |\lambda|$) is much larger than spatial variation of $m_0$ and $\lambda$ with $y$, the transition between $\mathcal{C}=0$ phase and $\mathcal{C}=1$ phase is adiabatic over $y$. As a result, the momentum $k_x$ and $k_y$ are good quantum numbers and the Hamiltonian [Eq.~\eqref{microH}] can be written as
\begin{equation}
\mathcal{H}(\mathbf{k})=-vk_x\sigma_y\zeta_z+vk_y\sigma_x\zeta_z+m_0\zeta_x+\lambda\sigma_z,
\end{equation}
where the low-energy wave functions of the Hamiltonian at $k_{x,y}=0$ near the transition point ($|m_0| \approx |\lambda|$) are given by $\psi_1=1/\sqrt{2}(0,0,1,1)^T$ and $\psi_2=1/\sqrt{2}(-1,1,0,0)^T$ with energies $E_1=|m_0|-|\lambda|$ and $E_2=-|m_0|+|\lambda|$, respectively. Projecting the Hamiltonian [Eq.~\eqref{microH}] into this low-energy subspace ($\psi_{1,2}$), we have the Hamiltonian near the transition point ($|m_0| \approx |\lambda|$) as
\begin{equation}\label{eq:LowEH}
\mathcal{H}(x)=-\widetilde{m}(y)\rho_z-v p_y\rho_x - v p_x\rho_y,
\end{equation}
where $\widetilde{m}(y)=|\lambda(y)|-|m_0(y)|$ and $\rho $ is the Pauli matrix acting in the $\psi_{1,2}$ subspace. Since our aim is to derive the effective $p$-wave pairing potential for the edge state which lives at the boundary between the $\mathcal{C} = 0$ and $\mathcal{C} = 1$ domain, we consider the quantity $\widetilde{m}(y)$ to change sign at the boundary ($y = 0$) where the variation is slow compared to the gap of the upper bands ($|m_0| + |\lambda|$) and fast relative to the gap of the lower bands ($||m_0|  -|\lambda||$). We take
\begin{equation}
\widetilde{m}(y)=\begin{cases}
-\widetilde{m}_1,& \text{for $y<0$,}\\
\widetilde{m}_2,& \text{for $y>0$,}
\end{cases}
\end{equation}
with $\widetilde{m}_1, \widetilde{m}_2>0$. In the following, we focus on $k_x \approx 0$ where the Hamiltonian at $k_x = 0$ can be written as
\begin{equation}
\textit{H}_0(y)=-\widetilde{m}(y)\rho_z+iv\partial_y\rho_x.
\label{H}
\end{equation}
The above Hamiltonian admits a zero-energy Majorana mode at the boundary due to the fact that $\widetilde{m}$ changes sign at the boundary. The Schr\"{o}dinger equation $H_0(y)\phi(y) = 0$ for the Majorana mode can be written as
\begin{equation}
\partial_y\phi(y) = -\frac{\widetilde{m}(y)}{v}\rho_x \phi(y).
\end{equation}
Solving the above equation, we obtain~\cite{PhysRevD.13.3398,shen2012topological}
\begin{align}\label{phi}
\phi_{\pm}(y)&=\sqrt{\frac{\widetilde{m}_1 \widetilde{m}_2}{v(\widetilde{m}_1+\widetilde{m}_2)}}\begin{pmatrix} 1 \\ \pm i \end{pmatrix}\exp\left(\mp\int^y_0\frac{\widetilde{m}(y')}{v} dy'\right),
\end{align}
where $\phi_{+} (\phi_{-})$ is the solution for $y >0(y <0)$.

We note that the BdG wave function is given by $\Psi_{k_x}(y)=\begin{pmatrix} \psi_{k_x}(y) \\ \bar{\psi}_{k_x}(y) \end{pmatrix}$ where the particle- and hole-component of the wave function are related by $\bar{\psi}_{k_x}(y)=i\sigma_y\psi^*_{k_x}(y)$. The superconducting term can be written in terms of this BdG wavefunction as
\begin{align}\label{eq:delta}
\widetilde{\Delta} &= \Delta_0\int^\infty_{-\infty} dy \Psi_{-k_x}^\dagger(y) \tau_x \Psi_{k_x}(y), \nonumber\\
&=\Delta_0\int^\infty_{-\infty} dy \left[\psi^\dagger_{-k_x}(y) \bar{\psi}_{k_x}(y) + \bar{\psi}^\dagger_{-k_x}(y)\psi_{k_x}(y)\right], \nonumber\\
&=\Delta_0\int^\infty_{-\infty} dy \left[ \psi^\dagger_{-k_x}(y)(i\sigma_y)\psi^*_{k_x}(y)+\psi^T_{-k_x}(y)(-i\sigma_y)\psi_{k_x}(y)\right], \nonumber\\
&=\Delta_0\int^\infty_{-\infty} dy \left(\left[\psi^T_{-k_x}(y)(i\sigma_y)\psi_{k_x}(y)\right]^*-\left[\psi^T_{-k_x}(y)(i\sigma_y)\psi_{k_x}(y)\right]\right),\nonumber\\
&=-2\Delta_0\mathrm{Im}\left[\int^\infty_{-\infty} dy\psi^T_{-k_x}(y)(i\sigma_y)\psi_{k_x}(y)\right],
\end{align}
where $\Delta_0$ is the proximity-induced $s$-wave pairing potential. To evaluate Eq.~\eqref{eq:delta}, we first expand the wave function near $k_x = 0$ using the first-order perturbation theory with $-v k_x \rho_y$ being the perturbation term. The wave function is given by
\begin{equation}
\psi_{k_x}(y)=\phi(y)+vk_x\phi'(y).
\end{equation}
where $\phi' = \left.\frac{1}{v}\frac{d\psi}{d k_x}\right|_{k_x = 0}$.
Plugging this into Eq.~\eqref{eq:delta}, we have
\begin{equation}
\widetilde{\Delta}=-2\Delta_0 \mathrm{Im}\left(\int^\infty_{-\infty} dy \left[ \phi^T(y)-vk_x\phi^{'T}(y)\right](i\sigma_y)\left[\phi(y)+vk_x\phi^{'}(y)\right]\right).
\end{equation}
Since $[\phi^{'T}(x)\sigma_y\phi(x)]^T = -\phi^T(y)\sigma_y\phi'(y)$ and $u^T\sigma_yu = 0$ due to the fact that $\sigma_y$ is antisymmetric, we then have
\begin{equation}
\Delta=-4v\Delta_0\mathrm{Im}\left(\int^\infty_{-\infty} dy\phi^T(y)(i\sigma_y)\phi'(y)\right),
\label{Deltap}
\end{equation}
where $\Delta = \widetilde{\Delta}/k_x$ is the effective $p$-wave pairing potential.
To evaluate the integral in Eq.~\eqref{Deltap}, we first calculate the first-order perturbed wave function $\phi'(y)$ which is given by
\begin{equation}
\phi'(y)=-\int^\infty_{-\infty}G(y,y')\rho_y\phi(y')dy',
\label{phi'}
\end{equation}
where the Green's function satisfy
\begin{equation}\label{eq:greenfunction}
\left(\widetilde{m}(y)\rho_y+v\partial_y\right)G(y, y')=i\rho_x\delta(y-y').
\end{equation}
We solve Eq.~\eqref{eq:greenfunction} by first considering $y'<0$, where we obtain
\begin{subequations}\label{eq:greenfunction1}
\begin{align}
G(y, y')&=\exp\left(\int^y_{y'-\epsilon}-\frac{\widetilde{m}(y'')}{v}\rho_ydy''\right)G(y'-\epsilon, y'), &  \text{for $y<y'<0$}, \label{eq:G1}\\
vG(y'+\epsilon, y')&=vG(y'-\epsilon, y')+i\rho_x, &  \text{for $y=y'<0$},\label{boundary11}\\
G(y, y')&=\exp\left(\int^y_{y'+\epsilon}-\frac{\widetilde{m}(y'')}{v}\rho_ydy''\right)G(y'+\epsilon, y'), & \text{for $y'<y<0$}\label{boundary12},\\
G(y, y')&=\exp\left(\int^y_{0}-\frac{\widetilde{m}(y'')}{v}\rho_ydy''\right)G(0,y'), & \text{for $y'<0<y$}\label{eq:G4}
\end{align}
\end{subequations}
Since the Green's function vanishes for $y \rightarrow \pm \infty$, Eqs.~\eqref{eq:G1} and Eqs.~\eqref{eq:G4} imply that $G(y'-\epsilon, y')$ and $G(0,y')$ must be the left eigenmatrix of $\rho_y$ with eigenvalue -1 and 1, respectively, i.e., $G(y'-\epsilon, y')= \begin{pmatrix} 1 \\ -i \end{pmatrix}\begin{pmatrix} a & b \end{pmatrix}$ and $G(0,y') = \begin{pmatrix} 1 \\ i \end{pmatrix}\begin{pmatrix} c & d \end{pmatrix}$.
Matching the Green's functions in Eq.~\eqref{eq:greenfunction1} for $y'<0$, we have
\begin{eqnarray}\label{eq:gplus}
G(y,y')=\left\{
  \begin{aligned}
G_{-,1}(y,y')=&\frac{1}{2v}\exp\left(\int^y_{y'-\epsilon}-\frac{\widetilde{m}_1}{v}dy''\right)\begin{pmatrix} 1 & -i \\ -i & -1\end{pmatrix},  &\text{for $y<y'<0$}, \label{G+1}\\
G_{-,2}(y,y')=&\frac{1}{2v}\exp\left(\int^y_{y'+\epsilon}-\frac{\widetilde{m}_1}{v}dy''\right)\begin{pmatrix} 1 & i \\ i & -1\end{pmatrix}, & \text{for $y'<y<0$}, \label{G+2}\\
G_{-,3}(y,y')=&\frac{1}{2v}\exp\left(\int^y_{0}-\frac{\widetilde{m}_2}{v}dy''\right)\exp\left(\frac{\widetilde{m}_1}{v}y'\right)\begin{pmatrix} 1 & i \\ i & -1\end{pmatrix}, & \text{for $y'<0<y$}. \label{G+3}
  \end{aligned}
  \right.
\end{eqnarray}
For the case where $y'>0$, we solve Eq.~\eqref{eq:greenfunction} and obtain
\begin{subequations}\label{eq:Gyplus}
\begin{align}
G(y, y')&=\exp\left(\int^y_{0}-\frac{\widetilde{m}(y'')}{v}\rho_ydy''\right)G(0, y'), &  \text{for $y\leq0<y'$}, \\
G(y, y')&=\exp\left(\int^y_{y'-\epsilon}-\frac{\widetilde{m}(y'')}{v}\rho_ydy''\right)G(y'-\epsilon, y'), & \text{for $0<y<y'$}\label{boundary21},\\
vG(y'+\epsilon, y')&=vG(y'-\epsilon, y')+i\rho_x, &  \text{for $y=y'>0$},\label{boundary22}\\
G(y, y')&=\exp\left(\int^y_{y'+\epsilon}-\frac{\widetilde{m}(y'')}{v}\rho_ydy''\right)G(y'+\epsilon,y'), & \text{for $0<y'<y$}.
\end{align}
\end{subequations}
By matching the boundary condition for the Green's functions in Eq.~\eqref{eq:Gyplus}, we then have
\begin{eqnarray}\label{eq:gmin}
G(y,y')=\left\{
  \begin{aligned}
G_{+,1}(y,y')=&\frac{1}{2v}\exp\left(\int^y_{0}-\frac{\widetilde{m}_1}{v}dy''\right)\exp\left(-\frac{\widetilde{m}_2}{v}y'\right)\begin{pmatrix} 1 & -i \\ -i & -1\end{pmatrix},  &\text{for $y<0<y'$}, \label{G-1}\\
G_{+,2}(y,y')=&\frac{1}{2v}\exp\left(\int^y_{y'-\epsilon}-\frac{\widetilde{m}_2}{v}dy''\right)\begin{pmatrix} 1 & -i \\ -i & -1\end{pmatrix}, & \text{for $0<y<y'$}, \label{G-2}\\
G_{+,3}(y,y')=&\frac{1}{2v}\exp\left(\int^y_{y'+\epsilon}-\frac{\widetilde{m}_2}{v}dy''\right)\begin{pmatrix} 1 & i \\ i & -1\end{pmatrix}, & \text{for $0<y'<y$}. \label{G-3}
\end{aligned}
\right.
\end{eqnarray}
Plugging in Eqs.~\eqref{eq:gplus} and ~\eqref{eq:gmin} into Eq.~\eqref{eq:greenfunction}, we can solve for $\phi'(y)$. For $y<0$, we have
\begin{eqnarray}
\begin{aligned}
\phi'_-(y)&=\int^{y-\epsilon}_{-\infty}G_{-,2}(y,y')\rho_y\phi_{-}(y')dy'+\int^{0}_{y+\epsilon}G_{-,1}(y,y')\rho_y\phi_{-}(y')dy'+\int^{\infty}_{0}G_{+,1}(y,y')\rho_y\phi_{+}(y')dy',
\\
&=\frac{1}{2}\sqrt{\frac{\widetilde{m}_1 \widetilde{m}_2}{v (\widetilde{m}_1+\widetilde{m}_2)}}\exp\left(\frac{\widetilde{m}_1}{v}y\right)\left[\frac{1}{\widetilde{m}_1}\begin{pmatrix} 1 \\ i \end{pmatrix}-\frac{1}{\widetilde{m}_2}\begin{pmatrix} 1 \\ -i \end{pmatrix}\right],
\end{aligned}
\label{phi'-}
\end{eqnarray}
and for $y>0$, we obtain
\begin{eqnarray}
\begin{aligned}
\phi'_+(y)&=\int^{0}_{-\infty}G_{-,3}(y,y')\rho_y\phi_{-}(y')dy'+\int^{y-\epsilon}_{0}G_{+,3}(y,y')\rho_y\phi_{+}(y')dy'+\int^{\infty}_{y+\epsilon}G_{+,2}(y,y')\rho_y\phi_{+}(y')dy',
\\
&=\frac{1}{2}\sqrt{\frac{\widetilde{m}_1 \widetilde{m}_2}{v (\widetilde{m}_1+\widetilde{m}_2)}}\exp\left(-\frac{\widetilde{m}_2}{v}y\right)\left[\frac{1}{\widetilde{m}_1}\begin{pmatrix} 1 \\ i \end{pmatrix}-\frac{1}{\widetilde{m}_2}\begin{pmatrix} 1 \\ -i \end{pmatrix}\right].
\end{aligned}
\label{phi'+}
\end{eqnarray}
Finally, using Eqs.~\ref{phi},~\eqref{phi'-}, and \eqref{phi'+} in Eq.~\eqref{Deltap}, we can write
\begin{eqnarray}
\begin{aligned}
\Delta&=-4v\Delta_0 \mathrm{Im}\left[\int^0_{-\infty}\phi_-(y)^T(i\sigma_y)\phi'_-(y)dy+\int^{\infty}_{0}\phi_+(y)^T(i\sigma_y)\phi'_+(y)dy\right],\\
&=-4\Delta_0\frac{1}{(\widetilde{m}_1+\widetilde{m}_2)} \left[\int^0_{-\infty} \widetilde{m}_2\exp\bigg(2\frac{\widetilde{m}_1}{v}y\bigg)dy+\int^{\infty}_{0} \widetilde{m}_1\exp\bigg(-2\frac{\widetilde{m}_2}{v}y\bigg)dy\right],\\
&=-\frac{2v\Delta_0}{(\widetilde{m}_1+\widetilde{m}_2)}\left(\frac{\widetilde{m}_2}{\widetilde{m}_1}+\frac{\widetilde{m}_1}{\widetilde{m}_2}\right).
\end{aligned}
\end{eqnarray}
Since $\widetilde{m}_{1,2}\sim\lambda$, we have
\begin{equation}
\Delta\sim v\Delta_0/\lambda.
\end{equation}

\section{Dephasing model of  chiral edges in a disordered superconducting system}
\begin{figure}[h!]
\centering
\includegraphics[width=0.3\linewidth]{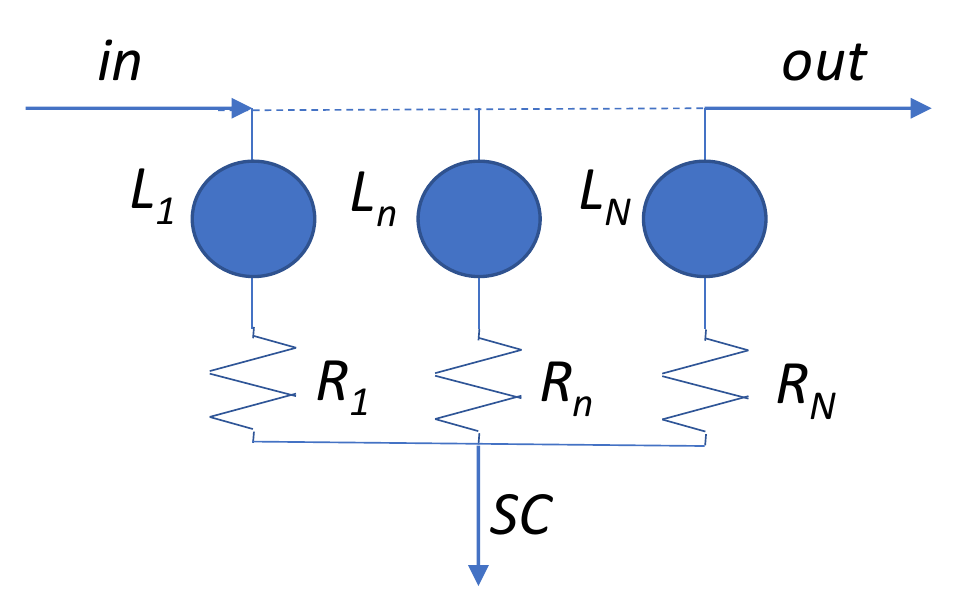}
\caption{\label{fig:network_model} Dephasing model of  chiral edges in a disordered superconducting system. The dephasing occur through tunneling processes from the chiral edges (arrowed lines) into the disjointed chiral loops (shown by blue discs and labeled by $L_n$ where $n = 1\cdots N$). The coupling between the loop $L_n$ and SC is assumed to be sufficiently weak such that the Cooper pairs are incoherently transferred through a resistance $R_n$ between them. The loop $L_n$ is assumed to be a reservoir in equilibrium at voltage $v_n$ relative to the SC.}
\end{figure}

To model the quantum Hall transition region we consider the chiral edge states to be coupled by tunneling through a series of loops (shown in Fig.~\ref{fig:network_model}). The coupling between the loops allows transfer of charges between the edges leading to a finite longitudinal conductance and non-quantized Hall conductance in the transition region. We further assume that the tunneling between the loops is smaller than the relaxation rate of electrons so that the loops may be considered to be reservoirs with potential $v_n$. More specifically, the incoherent chiral edge may be considered
of as a multiterminal system with leads at the \textit{in} and \textit{out}
ends as well as at each of the loops. The proximity coupling to the superconductor leads to a finite resistance $R_n$ connecting the loops to the superconductor allowing the loops to drain charge into the superconductor which is at zero voltage.

The outgoing current $i_{\mathrm{out}}$ is given by
\begin{align}\label{eq:iout}
i_{\mathrm{out}} &= (V_{\mathrm{in}} - V_3)g_{\mathrm{trans}} = \Lambda_{\mathrm{in},\mathrm{out}} (V_{\mathrm{in}} - V_3) + \sum_{n} \Lambda_{n,\mathrm{out}}v_n,
\end{align}
where
\begin{align}
 v_n &= \text{ the voltage at loop } L_n,\nonumber\\
\Lambda_{\mathrm{in},\mathrm{out}} &= \text{ the response of the outgoing current}  \text{ to the
voltage in lead } in,\nonumber\\
\Lambda_{n,\mathrm{out}} &= \text{ the response of the outgoing current}  \text{ to the
voltage in lead } n,\nonumber\\
g_{\mathrm{trans}} &= \text{ the transconductance due to the chiral edge modes.}
\end{align}
From Eq.~\eqref{eq:iout}, we obtain the transconductance to be
\begin{align}
&g_{\mathrm{trans}}=\Lambda_{\mathrm{in},\mathrm{out}}+\sum_n \Lambda_{n,\mathrm{out}} \frac{v_n}{V_{\mathrm{in}}-V_3}.\label{eqgchs}
\end{align}
The current flowing through loop $L_n$ is
\begin{align}\label{eq:in}
i_n = (V_{\mathrm{in}}-V_3)\Lambda_{\mathrm{in},n} + \sum_{m<n} \Lambda_{mn}v_m =v_n\left[R_n^{-1} +  \Lambda_{n,\mathrm{out}}+\sum_{m>n} \Lambda_{nm}  \right],
\end{align}
where
\begin{align}
 v_n &= \text{ the voltage at loop } L_n,\nonumber\\
\Lambda_{\mathrm{in},n} &=  \text{ the response of the current in lead } n \text{ to the
voltage in lead } in,\nonumber\\
\Lambda_{n,\mathrm{out}} &=  \text{ the response of the outgoing current}  \text{ to the
voltage in lead } n,\nonumber\\
\Lambda_{mn} &=  \text{ the response of the current in lead } n \text{ to the
voltage in lead } m,\nonumber\\
R_n &= \text{ the resistance between loop } L_n \text{ and SC}.
\end{align}
From Eq.~\eqref{eq:in}, we then obtain the recursion relation for the voltages $v_n$ at loop $L_n$ as
\begin{align}
&\frac{v_n}{V_{\mathrm{in}}-V_3}=\frac{\Lambda_{\mathrm{in},n}+\sum_{m<n}\Lambda_{mn}\frac{v_m}{V_{\mathrm{in}}-V_3}}{R_n^{-1}+\Lambda_{n,\mathrm{out}}+\sum_{m>n}\Lambda_{nm}}.
\end{align}

\section{Transmission Matrix}\label{sec:scatmat}
Using the BdG Hamiltonian in Eq.~\eqref{model}, we can write the Schr\"{o}dinger equation $\mathcal{H}_{\mathrm{BdG}}\psi=E\psi$ for the edge state as
\begin{equation}\label{hamil}
-i\tilde{v}(x)\partial_x\psi_E(x)=\left(E\tau_0+\mu(x)\tau_z+\frac{1}{2}i\partial_x\tilde{v}(x)\right)\psi_E(x),
\end{equation}
where $\tilde{v}(x)=v\tau_0+\Delta(x)\tau_x$ is the effective edge mode velocity.

\begin{figure}[h!]
\centering
\includegraphics[width=0.3\linewidth]{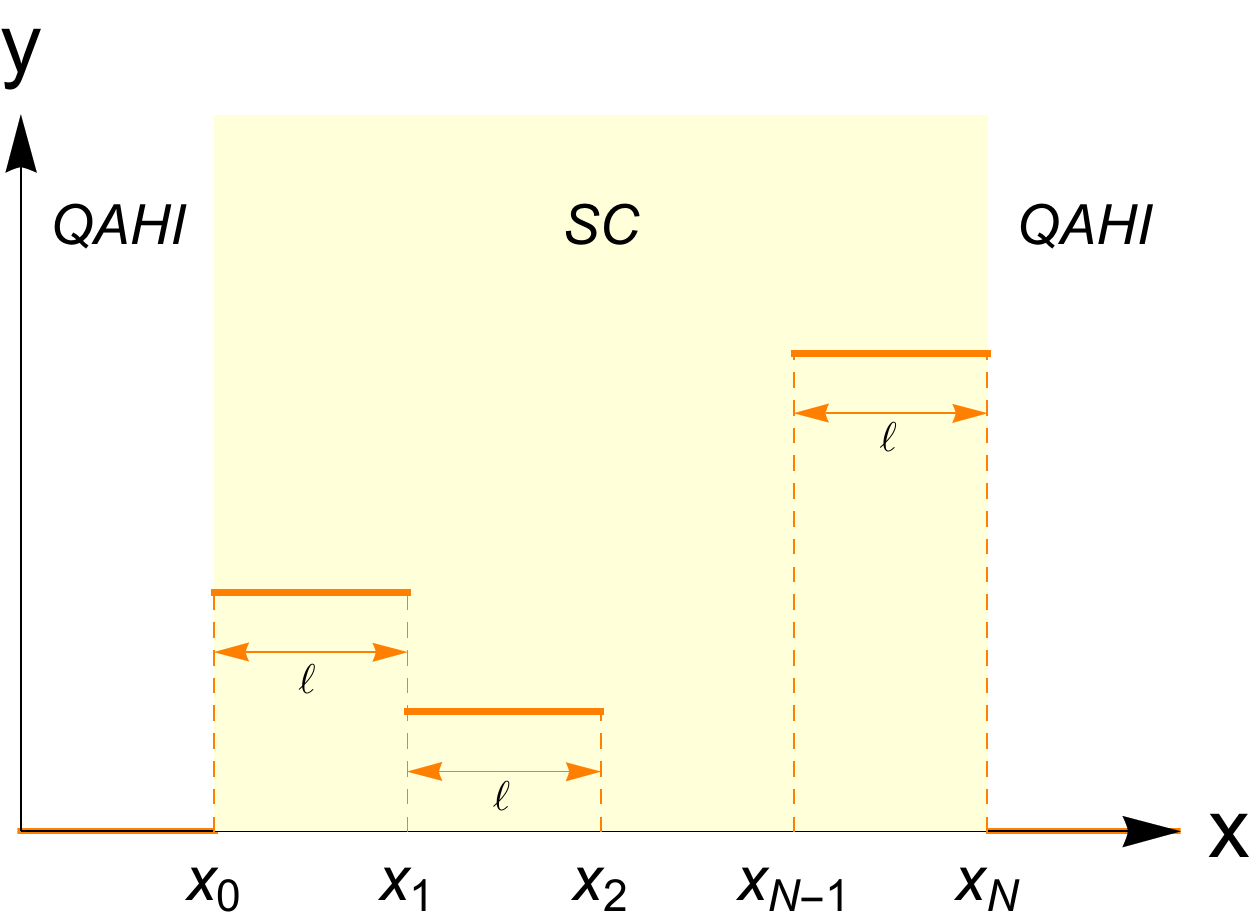}
\caption{\label{SFig1} Spatial variation of $y = \mu, \Delta$ along the electron trajectory $L$. The lattice length is $\ell$. For the numerical simulation done in this paper, we set $\ell = 0.05\overline{\eta}$.}
\end{figure}
 As shown in Fig.~\ref{SFig1}, we assume $\tilde{v}(x)$ to be spatially varying along the electron trajectory $L$ in the SC region where $\tilde{v}(x)$ jumps at the boundaries (at $x=x_0,\cdots x_N$). Since $\partial_x{\tilde{v}}(x)\gg E,\mu$ at the boundaries, integrating Eq.~\eqref{hamil} at the boundaries yields
\begin{equation}
\psi_E(x_{j+})=\frac{\tilde{v}_{j}^{1/2}}{\tilde{v}_{j+1}^{1/2}}\psi_E(x_{j-}).
\end{equation}
Here, $x_{j\pm}$ is the position just to the right/left of the boundary $j$. In the region between $x_{j-1}$ and $x_{j}$, $\tilde{v}(x) = \mathrm{const}$, so we have
\begin{equation}\label{eq:vjconst}
\psi_E(x_{j-})=e^{i\tilde{v}_{j}^{-1}(\mu_j\tau_z+E\tau_0)\ell}\psi_E(x_{(j-1)+}).
\end{equation}
Using the Baker-Campbell-Hausdorff formula $e^{iA^{-1}B}=A^{-1/2}e^{i A^{-1/2}BA^{-1/2}}A^{1/2}$, we can write down Eq.~\eqref{eq:vjconst} as
\begin{equation}
\psi_E(x_{j-})=\tilde{v}_{j}^{-1/2}e^{i\tilde{v}_{j}^{-1/2}(\mu_j\tau_z+E\tau_0)\tilde{v}_{j}^{-1/2}\ell}\tilde{v}_{j}^{1/2}\psi_E(x_{(j-1)+}).
\end{equation}
The outgoing current amplitude at $x_N = L$ is then given by
\begin{equation}
\mathcal{J}_{\mathrm{out}}(E)=\prod_je^{i\tilde{v}_{i}^{-1/2}(\mu_j\tau_z+E\tau_0)\tilde{v}_{j}^{-1/2}\ell}\mathcal{J}_{\mathrm{in}}(E),
\label{finalstate}
\end{equation}
where $\mathcal{J}_{\mathrm{out}}(E) = \sqrt{v}\psi_E(x_N)$ and $\mathcal{J}_{\mathrm{in}}(E) = \sqrt{v}\psi_E(x_0)$ are the outgoing and incoming current amplitudes in the SC region. The transmission matrix along the electron trajectory with spatially varying $\Delta$ and $\mu$ is given by
\begin{equation}
\mathcal{\widetilde{T}}(E) = \prod_je^{i\tilde{v}_{i}^{-1/2}(\mu_j\tau_z+E\tau_0)\tilde{v}_{j}^{-1/2}\ell}.
\end{equation}
Since the transmission matrix in the QAHI region gives just a trivial phase factor in the current amplitude, the length of QAHI region does not change the conductance and hence is ignored in the calculation. Furthermore, since each chiral loop $L_n$ in Fig.~\ref{fig:network_model} is assumed to be coupled to the chiral edges with strength $\Omega_n$, the transmission matrix between loop $L_m$ and $L_n$ from the chiral edge is given by
\begin{equation}\label{eq:scatmats}
\mathcal{T}_{mn}(E) =\zeta_{mn} \prod_{m<j<n} e^{i\tilde{v}_j^{-1/2}(\mu_j\tau_z+E\tau_0)\tilde{v}_j^{-1/2}\ell},
\end{equation}
where $\zeta_{mn}=\Omega_m\Omega_n\prod_{m<j<n}(1-\Omega_j^2)^{1/2}$ is a numerical factor which is related to the
couplings $|\Omega_j|<1$ of the chiral edge to the lead $j$ ($\Omega_{\mathrm{in}}=\Omega_{\mathrm{out}}\equiv 1$).

\section{Dependence of the plateau width on pairing potential }\label{sec:width}
\begin{figure}[h!]
\centering
\includegraphics[width=0.5\linewidth]{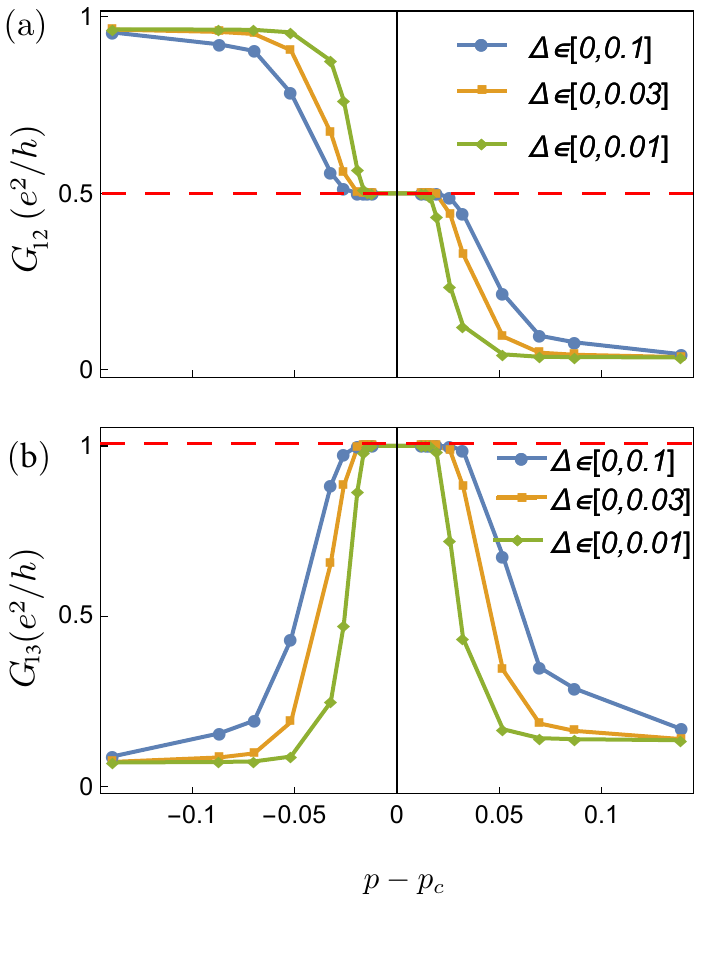}
\caption{\label{SFig2} Conductances (a) $G_{12}$ and (b) $G_{13}$ for different pairing potentials $\Delta$. The red dashed lines denote (a) $G = e^2/2h$  and (b) $G = e^2/h$. The parameters used here are the same as those used in Fig.~\ref{Fig4}. Note that the plateau width of $G_{12}$ and $G_{13}$ decrease with decreasing $\Delta$.}
\end{figure}
In the main text, we have calculated the conductances $G_{12}$ and $G_{13}$ using Eqs.~(\ref{G12}) and ~(\ref{G13}). As shown in Fig.~\ref{Fig4}, $G_{12}$ exhibits a half-quantized plateau and $G_{13}$ exhibits an integer quantized plateau. In this section, we compute the conductances for different values of pairing potential $\Delta$ for the case where there is quasiparticle leakage from the edge state into the adjacent chiral loops.

The plots of the conductance $G_{12}$ and $G_{13}$ for different pairing potentials $\Delta$ are shown in Fig.~\ref{SFig2}(a) and (b). As seen from the figures, $G_{12}$ and $G_{13}$ are perfectly quantized at $e^2/2h$ and $e^2/h$, respectively, at $p=p_c$ regardless of the pairing potentials. The width of the conductance plateaus also decrease with decreasing pairing potential $\Delta$. This can be easily understood as Andreev scattering amplitudes becoming smaller for smaller pairing potential. As a result, more scattering processes are required to average out the Bogoliubov quasiparticle charge to zero. This implies that the quasiparticles need to be scattered through a longer trajectory length $L$ which in turn necessitates the system to be closer to the percolation threshold. In the case where there is no superconductivity, Andreev scatterings are absent. As a result, the conductance profiles do not develop any plateaus. We note that the above results also hold for the case where there is no dissipation.

\section{The dephasing model via random thermal fluctuations}\label{sec:nodissipation}

Instead of modeling the dephasing process via tunneling of the chiral edge into adjacent chiral loops, in this section, we model the dephasing effect microscopically as arising from random thermal fluctuations of the disorder potential due to the interaction with the motion of charge impurities~\cite{Feng1960Sensitivity,Lee1987Universal,PhysRevB.75.104202}. Since there is no dephasing into the chiral loops, there is no coupling between the chiral edge and the chiral loops $L_n$, i.e., $\Omega(x)=0$.  Accordingly, the transmission matrix $\mathcal{T}_{mn}$ is zero except $\mathcal{T}_{\mathrm{in},\mathrm{out}}$. With $\Lambda_{mn} = 0$ except $\Lambda_{\mathrm{in},\mathrm{out}}$, the chiral edge may be thought of as a two-terminal system with leads at the $in$ and $out$ end.  To prohibit the incoherent charge transfer between the loops $L_n$ and SC, the resistance $R_n$ between them is set to be very large. As a result, the transconductance in Eq.~\ref{eqgch} becomes $g_{\mathrm{trans}}=\Lambda_{\mathrm{in},\mathrm{out}}$.

\begin{figure}[h!]
\centering
\includegraphics[width=0.5\linewidth]{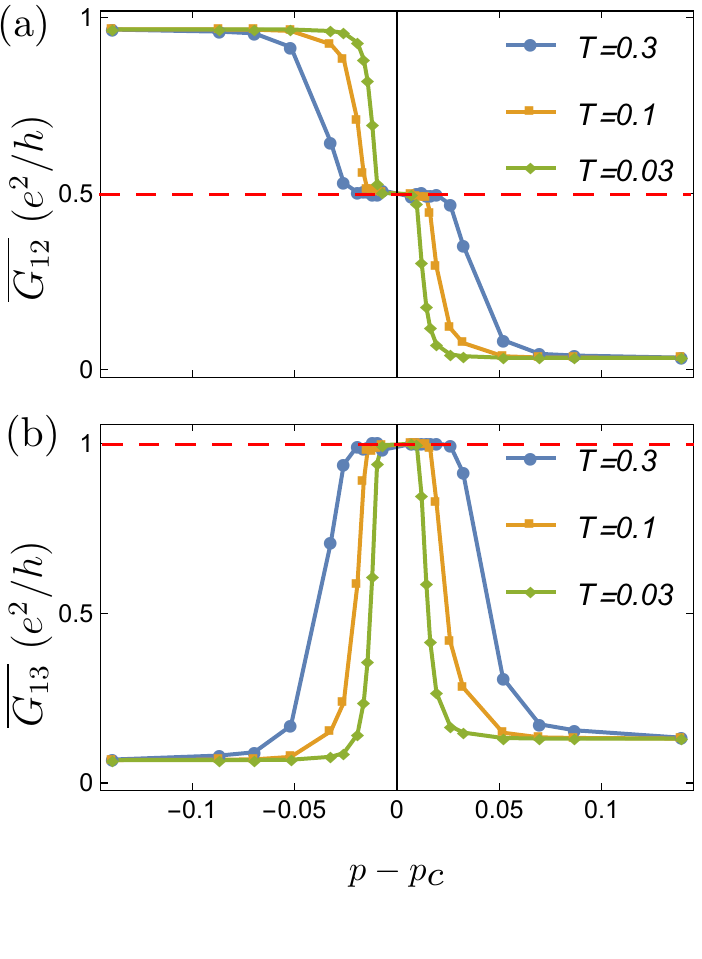}
\caption{\label{SFig3} Disorder-averaged conductances (a) $\overline{G_{12}}$ and (b) $\overline{G_{13}}$ for different temperature $T$ for the case with random thermal fluctations and no quasiparticle tunneling from the chiral edge into adjacent chiral loops [$\Omega(x)=0$ and $R(x)\rightarrow\infty$]. The disorder is modeled as $\mu(x) = \mu_{0}(x) + \mu_{T}(x)$ and $\Delta(x) = \Delta_{0}(x) + \Delta_{T}(x)$ where $\mu_{0} \in [-0.009,0.009]$ and $\Delta_{0} \in [0,0.09]$ are the temperature-independent parts while $\mu_{T} \in [-0.001,0.001]$ and $\Delta_{T} \in [0,0.01]$ are the temperature-dependent parts. The red dashed lines denote (a) $G = e^2/2h$  and (b) $G = e^2/h$.  Note that the plateau width of $\overline{G_{12}}$ and $\overline{G_{13}}$ decrease with decreasing temperature $T$ where it vanishes in the zero-temperature limit.}
\end{figure}

Ideally at zero temperature, single disorder realizations are expected to lead to mesoscopic fluctuations of conductance with changes of parameters~\cite{Altshuler1985fluctuations,Lee1985Universal}.
 However, finite temperature washes out this fluctuation through a combination of smearing of the incoming electrons' energies by the Fermi function and also dephasing. Here, the dephasing effect is modeled as random thermal fluctuations of the disorder potential due to the interaction with the motion of charge impurities~\cite{Feng1960Sensitivity,Lee1987Universal,PhysRevB.75.104202}. More specifically, we introduce $V_0(x)$ as the zero-temperature fluctuation of the disorder potential arising from impurities. This potential is taken to be random from site to site but it is held fixed between the different samples used in the simulation. In addition, we introduce a ``thermal fluctuation" $V_{T}(x)$ of the background potential arising from the motion of the charge impurities due to finite temperature. The thermal fluctuation changes on a time-scale of the order of temperature (i.e., $\hbar/k_B T \sim 1$ ns) which is much shorter than the measurement time ($\sim$ 1 ms), so that a single measurement averages over many random realizations of the potential $V_{T}(x)$. Therefore, each of our plots below are calculated using a disorder potential $V(x) = V_0(x) + V_{T}(x)$ where we use a single realization of the potential $V_0(x)$ but average over an ensemble of realizations for $V_T(x)$. The potential $V_0(x)$ has a variance which is independent of temperature but the variance of $V_T(x)$ is of the order of temperature.

The plots of the disorder-averaged conductance $\overline{G_{12}}$ and $\overline{G_{13}}$ for this case are shown in Fig.~\ref{SFig3}. We can see that $\overline{G_{12}}$ exhibits a half-quantized plateau and $\overline{G_{13}}$ exhibits an integer quantized plateau. This result is similar to the case with quasiparticle leakage into the adjacent chiral loops. The main difference between the two cases lies in the low-temperature limit. For the case without quasiparticle tunneling into adjacent chiral loops, the plateau width is strongly dependent on temperature where it keeps decreasing with lowering temperature as shown in Fig.~\ref{SFig3} with the plateau vanishing at zero temperature. On the other hand, for the case where there is a dissipation due to the leakage of quasiparticles from the edge states into the adjacent chiral loops (Fig.~\ref{Fig4}), the plateau width is stable to relatively low temperatures (determined by system size and tunnel barriers) where the plateau width does not change in the low-temperature regime. Ultimately, as temperature goes to zero, the dissipation of the chiral loop is eliminated and the plateau disappears but depending on parameters, this can easily occur well below experimentally realistic temperatures.

\end{document}